\documentclass[reprint,prl,twocolumn,showpacs,superscriptaddress,aps,longbibliography]{revtex4-2}

\usepackage[colorlinks=true,citecolor=blue]{hyperref} 
\usepackage{graphicx}
\usepackage{bm}
\usepackage{amsmath}
\usepackage{amssymb}

\DeclareGraphicsExtensions{.png,.jpg,.eps}
\usepackage{xcolor}

\newcommand{\beq}{\begin{equation}}
\newcommand{\eeq}{\end{equation}}

\usepackage{mleftright} 

\newcommand{\figref}[1]{\mbox{Fig.~\ref{#1}}}

\renewcommand{\eqref}[1]{\mbox{Eq.~(\ref{#1})}}

\newcommand{\figpanel}[2]{Fig.~\hyperref[#1]{\ref*{#1}(#2)}}
\newcommand{\figpanels}[3]{Fig.~\hyperref[#1]{\ref*{#1}(#2)-(#3)}}
\newcommand{\figpanelNoPrefix}[2]{\hyperref[#1]{\ref*{#1}(#2)}}

\usepackage{siunitx}

\begin{document}

\author{Zina Lippo}
\affiliation{Department of Physics, University of Helsinki, Finland}

\author{Elizabeth Louis Pereira}
\affiliation{Department of Applied Physics, Aalto University, 02150 Espoo, Finland}

\author{Jose L. Lado}
\affiliation{Department of Applied Physics, Aalto University, 02150 Espoo, Finland}

\author{Guangze Chen}
\email{guangze@chalmers.se}
\affiliation{Department of Microtechnology and Nanoscience, Chalmers University of Technology, 41296 G\"{o}teborg, Sweden}

\title{Topological zero modes and correlation pumping in an engineered Kondo lattice}

\begin{abstract}
Topological phases of matter provide a flexible platform to engineer unconventional quantum excitations in quantum materials.
Beyond single particle topological matter, in systems with strong quantum many-body correlations, many-body effects
can be the driving force for non-trivial topology. Here, we propose a one-dimensional engineered Kondo lattice where the emergence of topological excitations is driven by collective many-body Kondo physics. 
We first show the existence of topological zero modes in this system by solving the interacting model with tensor networks,
and demonstrate their robustness against disorder.
To unveil the origin of the topological zero modes, we analyze the associated periodic Anderson model showing that
it can be mapped to a topological non-Hermitian model, 
enabling rationalizing the origin of the topological zero modes. 
We finally show that the topological invariant of the many-body Kondo lattice can be computed with a correlation matrix pumping method
directly with the exact quantum many-body wavefunction. 
Our results provide a strategy to engineer topological Kondo insulators, 
highlighting quantum magnetism as a driving force in engineering topological matter.
\end{abstract}

\date{\today}

\maketitle
{\it{Introduction.}}
The engineering of topological phases of matter~\cite{RevModPhys.82.3045,RevModPhys.83.1057} has 
provided a highly successful strategy to create electronic excitations beyond those found in
conventional materials, 
including chiral~\cite{RevModPhys.95.011002}, helical~\cite{Maciejko2011} and Majorana states~\cite{Alicea2012}. 
The robustness of topological excitations to disorder renders them of interest for a variety of applications,
ranging from electronics~\cite{RevModPhys.82.3045,RevModPhys.83.1057,RevModPhys.82.1959}, spintronics~\cite{mejkal2018} to topological quantum computing~\cite{RevModPhys.80.1083}. 
Topological phases are often challenging to find in naturally occurring materials~\cite{Deng2020,Wu2018,Jiao2020},
which has motivated a variety of efforts to create them in artificially engineered systems~\cite{Andrei2021,RevModPhys.91.041001,RevModPhys.91.015006}.
In particular, a variety of strategies can be leveraged
to engineer
these states, by combining competing orders~\cite{Alicea2012,Dvir2023,Kezilebieke2020,Schneider2022},
using external driving~\cite{Jotzu2014,McIver2019,Rudner2020}, leveraging
coupling to the environment~\cite{RevModPhys.93.015005,Dai2023,Harari2018} or
engineering
many-body interactions~\cite{PhysRevResearch.3.013265,Liu2021,Serlin2020}.
Beyond single-particle topological matter~\cite{annurev:/content/journals/10.1146/annurev-conmatphys-031214-014740},
the engineering of quantum many-body effects 
may ultimately allow creating topological states
that have no single particle counterpart~\cite{Zeng2023,Lu2024}.

\begin{figure}[t!]
\center
\includegraphics[width=\linewidth]{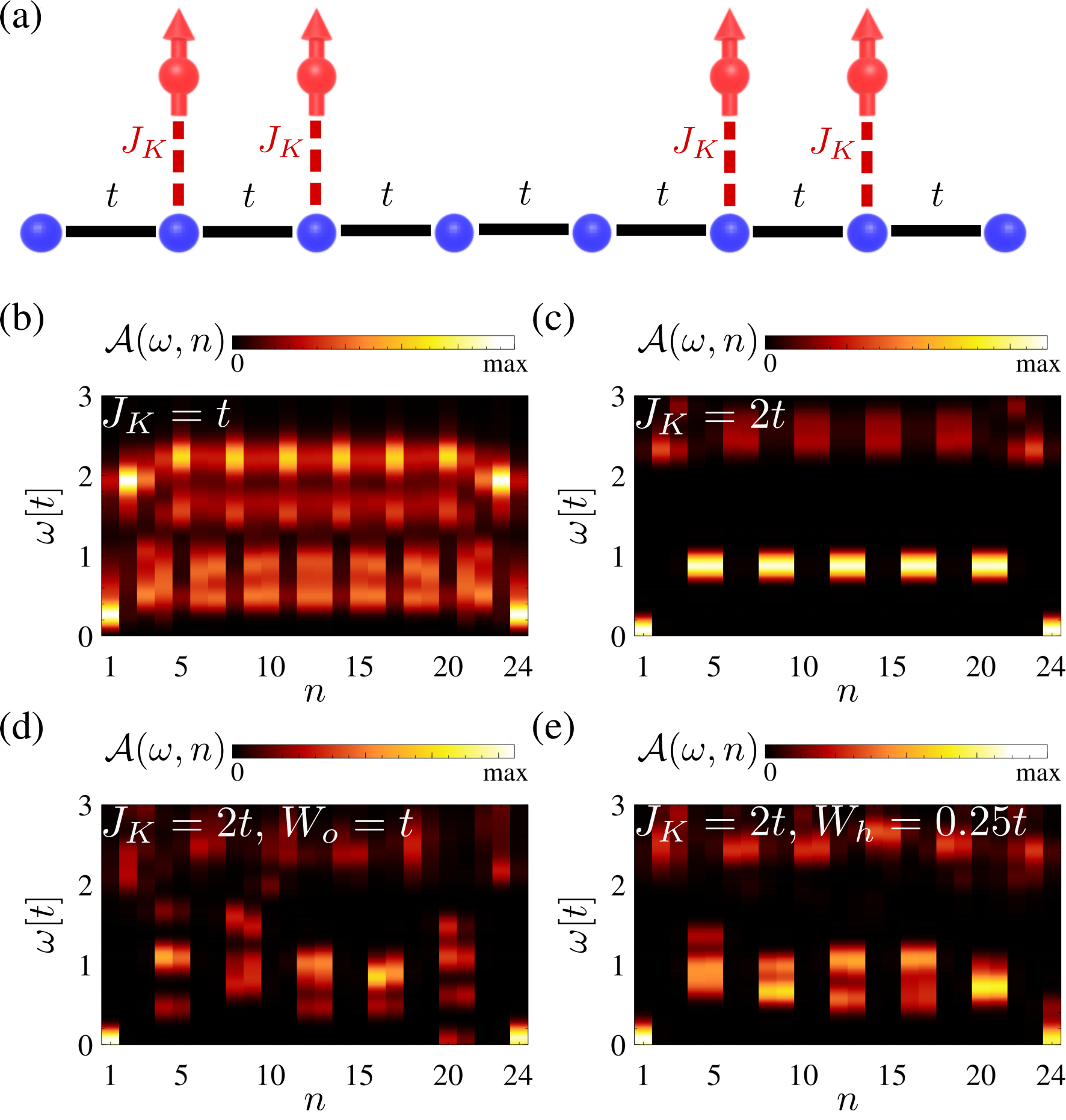}
\caption{
(a) Schematic
of the Kondo lattice model~\eqref{eq_model}, where the electronic sites (blue) have a uniform hopping $t$ between them. Kondo spins (red) are coupled to some of the electronic sites with strength $J_K$.
Panels (b,c) show the spectral function featuring the zero edge modes.
Panels (d,e) show the spectral function for a disordered system, demonstrating the robustness of the zero modes.
We took $J_K=t,2t$ in (b,c) and $J_K=2t$, $W_o=t$ and $W_h=0.25t$ in (d,e).
}
\label{fig1}
\end{figure}

Kondo lattices~\cite{RevModPhys.69.809} are a paradigmatic platform where many-body effects dictate
the interplay between electronic delocalization and magnetic entanglement formation~\cite{RevModPhys.56.755,Annurev_2016_Yazdani,Annurev_2016_Coleman,Wirth2016,Jiao2020,HF2021}. 
Topological states have been known to appear in topological
Kondo insulators~\cite{Dzero2016,PhysRevLett.104.106408},
where Kondo screening leads to an effectively
topological electronic structure for the single-particle excitations~\cite{Neupane2013,Xu2014}.
However, conventional mechanisms to create topological Kondo insulators require strong spin-orbit coupling of localized $f$ electrons~\cite{Dzero2016,PhysRevLett.104.106408}, and potential material candidates remain restricted~\cite{Xu2020,Pirie2023,Pirie2019,guerci2024topologicalkondosemimetalinsulator}. Thus, finding alternative strategies to engineer topological matter in Kondo systems will enable
creating topological excitations by using quantum magnetism as a fundamental driving force.

Here, we propose a design to realize a topological Kondo insulator in a one-dimensional Kondo
lattice, where the topology is solely generated by the many-body Kondo coupling. We first show the existence of zero edge modes
by exactly solving this system with tensor networks and demonstrate their robustness against disorder.
We then unveil the topological nature of the edge modes by performing a mapping to a periodic Anderson model,
and showing that the topological protection of these zero modes stems
from an effective non-Hermitian model. 
We finally show that the topological invariant can be exactly computed
with the many-body wavefunction using a correlation matrix pumping method,
that becomes equivalent to the well-known Zak phase for
non-interacting systems.

{\it{Model and Results.}} We consider the one-dimensional spin-1/2 Kondo lattice model of the form
\beq \label{eq_model}
\mathcal{H}= t\sum_{s,n=1}^{N-1}  \mleft( c_{n+1,s}^\dag c_{n,s} + h.c.\mright)
 +J_K\sum_{\substack{s,s',\alpha \\ n\in n_K}}  c_{n,s}^\dag \sigma^\alpha_{ss'} c_{n,s'} S^\alpha_{n}
\eeq
where $c_n^\dag$ and $c_n$ are electron creation and annihilation operators at site $n$, $\alpha=x,y,z$, $S^\alpha_n$ is the Kondo spin that is coupled to the electronic site $n$, $n_K=\{4m+1,4m+2|m\in\mathcal{N}\}$ is the collection of Kondo sites, and $t$ and $J_K$ are the hopping and Kondo coupling strengths~[\figpanel{fig1}{a}]. For the sake of concreteness, we consider a chain with $24$ fermionic sites and $12$ Kondo spin sites. 
We solve the many-body ground state of the system with a tensor-network formalism.
The many-body ground state is a non-magnetic spin singlet state, with all the Kondo
spin sites screened by the electronic gas, realizing a minimal
example of a non-uniform one-dimensional Kondo screened lattice.
The charge excitations of the many-body system can be obtained from the local electronic spectral function
\beq \label{eq_spec}
\mathcal{A}(\omega,n)= \langle \Omega|c_n \delta(\omega-\mathcal{H} + E_0) c^\dag_n|\Omega\rangle 
\eeq
where $|\Omega\rangle$ is the ground state of the system, $E_0$ is the ground state energy, and $\delta$
is the Dirac delta function.
The previous object can be computed with tensor networks using a Chebyshev algorithm~\cite{RevModPhys.78.275,PhysRevResearch.1.033009,itensor,DMRGpyLibrary}. 
As shown in \figpanels{fig1}{b}{c}, zero edge modes appear in the chain for non-zero $J_K$. Due to finite-size effect, these edge modes acquire a finite energy which decreases as $J_K$ increases. This shows that sufficiently large Kondo coupling can induce many-body zero mode excitations at the edge
in the Kondo lattice. The appearance of zero edge modes coexists
with gapped bulk electronic spectra, as expected from a topological state.

To show that these zero modes are topological, we now demonstrate their robustness against disorder. We consider onsite and
hopping disorder taking the form
$
\mathcal{H}_\text{onsite}= \sum_{s,n\neq1,N}W_o\chi_nc_{n,s}^\dag c_{n,s}
$
and 
$
\mathcal{H}_\text{hopping}= \sum_{s,n=1}^{N-1} W_h\chi_n\mleft(c_{n+1,s}^\dag c_{n,s}+h.c.\mright)
$
where $W_{o,h}$ are the onsite and hopping disorder strength, and $\chi_n$ are Gaussian distributed random variables with width 1. We consider $W_o=t$ and $W_h=0.25t$ in our case. Averaging over 5 different disorder configurations, we obtain \figpanels{fig1}{d}{e}, showing that the zero modes are robust against disorder.

\begin{figure}[t!]
\center
\includegraphics[width=\linewidth]{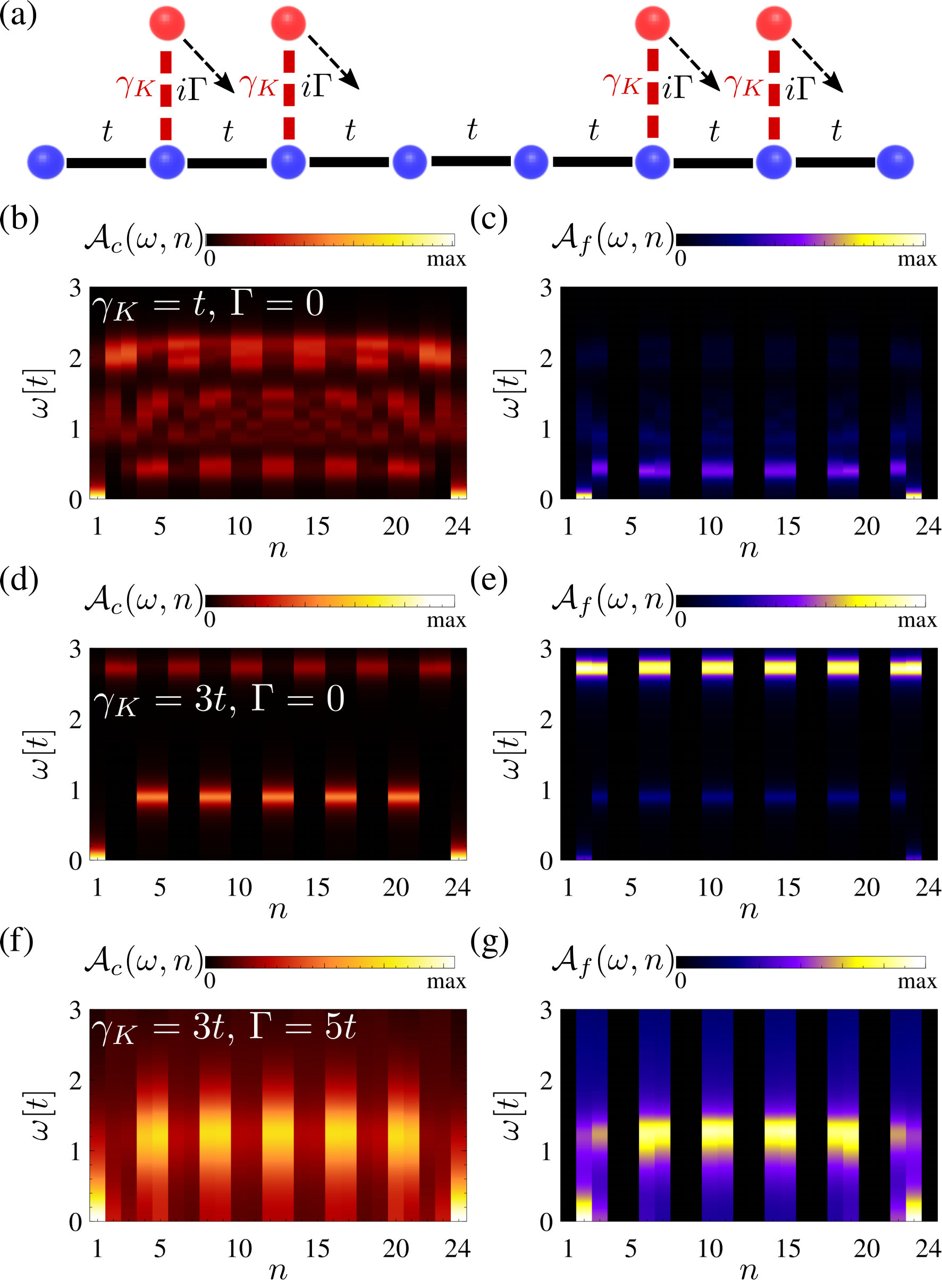}
\caption{Spectral function of the effective non-Hermitian Anderson model.
(a) The non-Hermitian effective Hamiltonian~\eqref{eq_MF} stemming from the Dyson equation, 
where localized fermions (red) are coupled to delocalized sites (blue) with coupling strength $\gamma_K$.
The spectral functions of the extended (b,d,f) and localized (c,e,g) fermions show the
existence of zero modes.
We took $\gamma_K=t$ and $\Gamma=0$ for (b,c), $\gamma_K=3t$ and $\Gamma=0$ for (d,e)
and $\gamma_K=3t$ and $\Gamma=5t$ for (f,g).
}
\label{fig2}
\end{figure}

\textit{Anderson lattice model.}
To rationalize the origin of the topological zero modes, we note that 
the Kondo lattice can be understood as stemming from
a periodic Anderson model~\cite{RevModPhys.69.809} of the form
\beq \label{eq_PAM}
\begin{aligned}
&H_\text{PAM}= t\sum_{s,n=1}^{N-1} \mleft(c_{n+1,s}^\dag c_{n,s}+h.c.\mright)\\
&+\gamma_K\sum_{s,n\in n_K} \mleft(c_{n,s}^\dag f_{n,s}+h.c.\mright)
+U\sum_{n\in n_K}f_{n,\uparrow}^\dag f_{n,\uparrow}f_{n,\downarrow}^\dag f_{n,\downarrow}
\end{aligned}
\eeq
where $f_{n,s}$ is the fermion on Kondo site $n$ with spin $s$. The localized fermions are coupled to the electrons with coupling strength $\gamma_K$, and they have an onsite interaction $U$. When $U$ is large, this model provides the same physics as the Kondo lattice model. We have neglected the small dispersion of the localized fermions.
We can now obtain an effective model for the delocalized electrons by including the interaction effects through a self-energy stemming
from a Dyson equation~\cite{PhysRevLett.125.227204}.
Due to the onsite interaction $U$, the localized fermions acquire a self-energy $\Sigma_f(\omega)=-a_1\omega-i(\Gamma+a_2\omega^2)$~\cite{PhysRevLett.125.227204} where $a_{1,2}$ and $\Gamma$ are coefficients. The frequency-dependent terms do not change the qualitative features of the electronic spectrum and are therefore neglected~\cite{PhysRevLett.125.227204}. This results in a finite quasiparticle lifetime $\tau=1/\Gamma$, and the inverse lifetime $\Gamma$ in general increases with temperature. With this treatment, we obtain an effective Hamiltonian for \eqref{eq_PAM} [\figpanel{fig2}{a}]:
\beq \label{eq_MF}
\begin{aligned}
&H_\text{eff}= t\sum_{s,n=1}^{N-1}\mleft(c_{n+1,s}^\dag c_{n,s}+h.c.\mright)\\ &+
\gamma_K\sum_{s,n\in n_K} \mleft(c_{n,s}^\dag f_{n,s}+h.c.\mright)
-i\Gamma\sum_{s,n\in n_K} f_{n,s}^\dag f_{n,s}
\end{aligned}
\eeq

To see the distribution of the zero modes after the hybridization, we compute the
spectral function of the extended and localized
fermions for $H_\text{eff}$ at different $\gamma_K$ [\figpanels{fig2}{b}{g}]. The
spectral functions are defined as
$\mathcal{A}_c(\omega,n)=\langle n|\delta(\omega-H_\text{eff})|n\rangle$
and $\mathcal{A}_f(\omega,n)=\langle n_f|\delta(\omega-H_\text{eff})|n_f\rangle$ where $|n\rangle$ is the local electronic basis at site $n$ and $|n_f\rangle$ is the localized fermion basis coupled to site $n$. We first consider the case $\Gamma=0$, i.e. the localized fermions have infinite lifetime. The zero edge modes appear for $\gamma_K>0$, with both extended and localized fermionic distributions [\figpanels{fig2}{b}{e}]. The ratio between the extended and localized fermionic parts of the zero modes depends on $\gamma_K$: the larger $\gamma_K$ is, the more distribution the zero modes have on the extended fermionic parts. 
Let us move on to consider $\Gamma>0$, i.e. the localized fermions have a finite lifetime, where we focus on $\gamma_K=3t$ regime and see how the finite lifetime influences the topological zero modes. Interestingly, the zero edge modes persist, and they have a lifetime much longer than $1/\Gamma$ [\figpanels{fig2}{f}{g}]. This shows the robustness of the topological zero modes against finite localized fermion lifetime.

\begin{figure}[t!]
\center
\includegraphics[width=\linewidth]{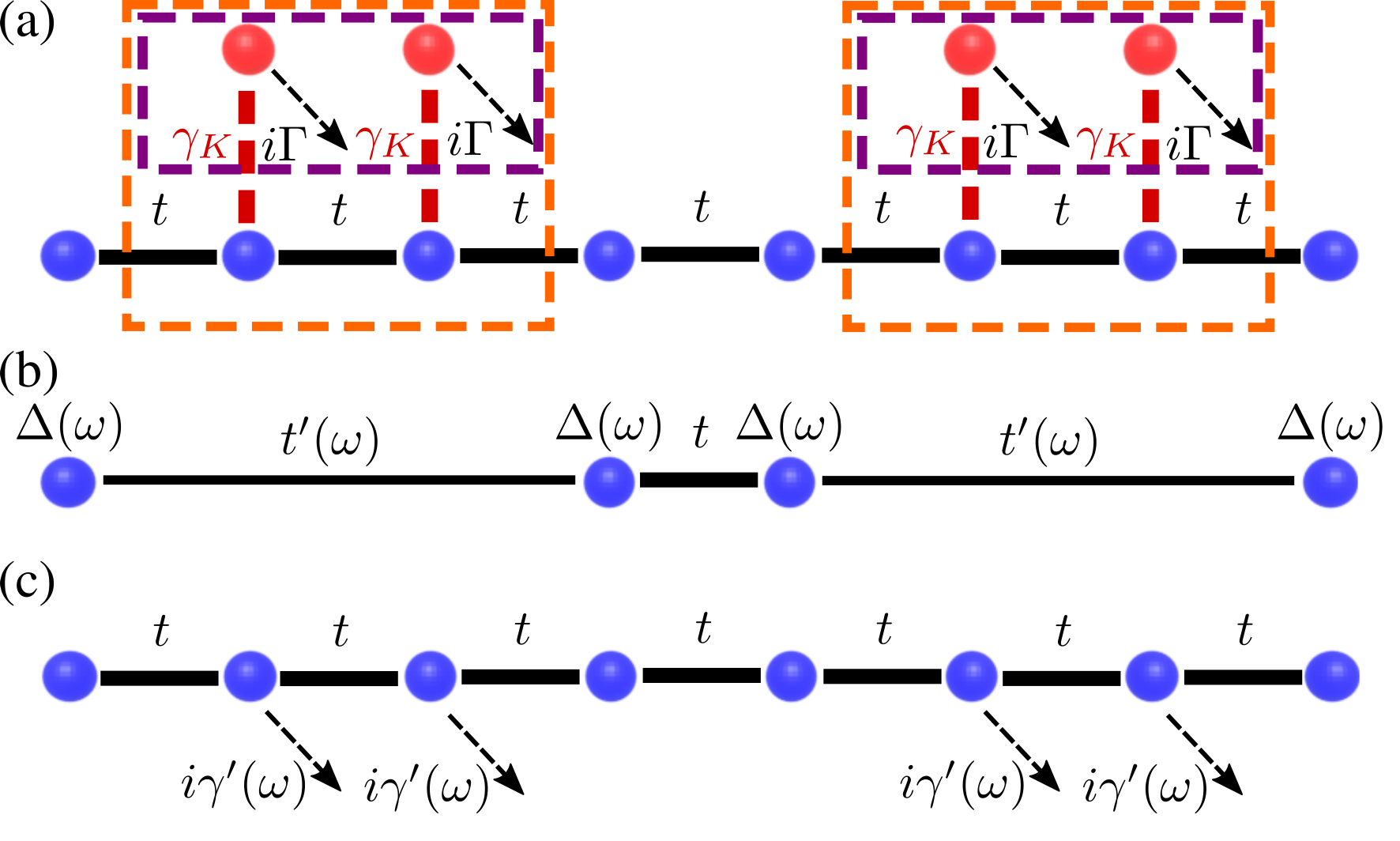}
\caption{Schematic of the strategies to derive the effective Hamiltonian.
Panel (a) shows two ways to separate the effective Hamiltonian \eqref{eq_MF} (dashed orange and purple rectangles).
Panel (b) shows the effective Hamiltonian obtained from the dashed orange separation in panel (a). This model has a frequency-dependent hopping $t'(\omega)$ and a uniform on-site loss $\Delta(\omega)$, given below \eqref{eq_H1eff}. At $\omega=0$, it reduces to an SSH model with $|t'|=t$, which is known to host topological zero modes.
Panel (c) shows the effective Hamiltonian obtained from the dashed purple separation in panel (a). This model has a frequency-dependent on-site loss $i\gamma'(\omega)$ given below \eqref{eq_H2eff}. At $\omega=0$, it reduces to a model with non-Hermitian topological zero edge modes.
}
\label{fig3}
\end{figure}

The origin of the topological zero modes in \figref{fig2} can be further clarified by integrating out the
interacting localized modes. This is done by separating the effective Hamiltonian $H_\text{eff}$ into two parts $H_1, H_2$ with coupling $H_{12}+H_{12}^\dag$, and tracing out the part $H_2$ to obtain an effective Hamiltonian $\mathcal{H}_\text{eff}(\omega)=H_1+\Sigma_e(\omega)$ where $\Sigma_e(\omega)=H_{12}^\dag(\omega-H_2)^{-1}H_{12}$ is the frequency-dependent self-energy. We first consider the following separation of the Hamiltonian (\figpanel{fig3}{a}): the non-Kondo coupled sites
$
H_1= t\sum_{s,n=1}^{N/4-1} c_{4n+1,s}^\dag c_{4n,s}+h.c. $,
the Kondo coupled sites
$
H_2= t\sum_{s,n=1}^{N/4-1} \mleft( c_{4n-2,s}^\dag c_{4n-1,s}+ h.c. \mright) +
\gamma_K\sum_{s,n\in n_K}  \mleft( c_{n,s}^\dag f_{n,s}+h.c. \mright)
-i\Gamma\sum_{s,n\in n_K}f_{n,s}^\dag f_{n,s}
$, and the coupling between both
$
H_{12}= t\sum_{s,n=1}^{N/4-1}\mleft(c_{4n-2,s}^\dag c_{4n-3,s}+c_{4n-1,s}^\dag c_{4n,s}\mright).
$
The effective Hamiltonian is then given by~\cite{footnote1}: [\figpanel{fig3}{b}]
\beq \label{eq_H1eff}
\begin{aligned}
\mathcal{H}_{\text{eff}}^{\text{SSH}} = & t\sum_{s,n=1}^{N/4-1}\mleft(c_{4n+1,s}^\dag c_{4n,s}+h.c.\mright)\\&+t'(\omega)\sum_{s,n=0}^{N/4-1} \mleft(c_{4n+4,s}^\dag c_{4n+1,s}+h.c.\mright)\\&+\Delta(\omega)\sum_{s,n=0}^{N/4-1} \mleft(c_{4n+1,s}^\dag c_{4n+1,s}+c_{4n+4,s}^\dag c_{4n+4,s}\mright)
\end{aligned}
\eeq
where
$
t'(\omega)=-t\frac{(\omega+i\Gamma)^2t^2}{(\omega+i\Gamma)^2t^2-\mleft[\omega(\omega+i\Gamma)-\gamma_k^2\mright]^2}\label{eq_t'}$
and $
\Delta(\omega)=-\frac{(\omega+i\Gamma)t^2\mleft[\omega(\omega+i\Gamma)-\gamma_k^2\mright]}{(\omega+i\Gamma)^2t^2-\mleft[\omega(\omega+i\Gamma)-\gamma_k^2\mright]^2}\label{eq_Delta}.
$
In particular, at $\omega=0$, $t'(0)=-t\frac{\Gamma^2t^2}{\Gamma^2t^2+\gamma_K^4}$ and $\Delta(0)=-it\frac{\Gamma t\gamma_K^2}{\Gamma^2 t^2+\gamma_K^4}$, and the effective Hamiltonian~\eqref{eq_H1eff} reduces to a topologically non-trivial Su-Schrieffer-Heeger (SSH) model with $|t'(0)|<t$ and a uniform onsite loss $\Delta(0)$. 

Beyond the topological origin based on the associated SSH model, 
the presence of loss 
motivates understanding the zero modes directly from the
non-Hermitian topology of the effective model.
This is obtained by the following separation
of $H_\text{eff}$, shown in \figpanel{fig3}{a}:
the delocalized sites
$ H_1= t\sum_{s,n=1}^{N-1}\mleft(c_{n+1,s}^\dag c_{n,s}+h.c.\mright)$,
the localized sites
$H_2=-i\Gamma\sum_{s,n\in n_K}f_{n,s}^\dag f_{n,s}$, and
the interaction between them
$
H_{12}= \gamma_K\sum_{s,n\in n_K} c_{n,s} f_{n,s}^\dag
$.
The effective Hamiltonian in this case is given by: [\figpanel{fig3}{c}] 
\begin{equation}\label{eq_H2eff}\mathcal{H}_{\text{eff}}^{\text{NH}} = t\sum_{s,n=1}^{N-1}\mleft(c_{n+1,s}^\dag c_{n,s}+h.c.\mright)
+i\gamma'_\omega \sum_{s, n\in n_K} c_{n,s}^\dag c_{n,s}
\end{equation}
where $\gamma'_\omega = -\frac{\gamma_K^2}{\Gamma-i\omega}$. At $\omega=0$, this model reduces to a non-Hermitian model known to be topological with zero modes for $\gamma_K^2/\Gamma\neq0$~\cite{PhysRevLett.121.213902,PhysRevB.100.161105,PhysRevLett.130.100401}. Thus, for $\Gamma\neq0$, the effective non-Hermitian model provides an alternative understanding of the topological zero modes.

{\it Many-body topological invariant.} The above analysis identifies the topological origin of the zero modes in the effective Hamiltonian~\eqref{eq_MF}. To be more concrete on the topological nature of the many-body zero modes, we introduce a correlation matrix pumping method to compute the many-body topological invariant of the Kondo lattice~\eqref{eq_model}. For a unit cell with twisted boundary conditions, the twist-dependent Hamiltonian is given by

\beq \label{eq_corr}
\begin{aligned}
\mathcal{H}^{\text{pump}}(\phi) =&t\sum_{s,n=1}^{N-1}\mleft(c_{n+1,s}^\dag c_{n,s}+h.c.\mright)
\\&+\sum_s \mleft(e^{i\phi}c^\dag_{N,s}c_{1,s}+h.c.\mright)
\\&+J_K\sum_{\substack{s,s',\alpha \\ n\in n_K}}c_{n,s}^\dag\sigma^\alpha_{ss'}c_{n,s'}S_{n}^\alpha
\end{aligned}
\eeq
The Hamiltonian has a twist-dependent ground state $|\Omega_\phi\rangle$, allowing us to define the correlation matrix~\cite{PhysRevB.89.085108,PhysRevB.90.085102,PhysRevB.92.075132,PhysRevB.109.195125} as
\begin{equation}
\Xi_{is,js'}(\phi) = \langle \Omega_\phi | c^\dagger_{is} c_{js'} | \Omega_\phi \rangle.
\end{equation}
This correlation matrix features eigenvectors $\Xi |v_\phi \rangle = \chi_\phi | v_\phi \rangle$, which in the non-interacting case directly correspond to the
single particle eigenstates of the Hamiltonian~\cite{PhysRev.97.1474,Siegbahn1981}. The eigenvalues $\chi_\phi$ of $\Xi$ are ranged in the interval $[0,1]$,
and in the non-interacting limit non-occupied eigenstates have eigenvalue $0$ whereas occupied states have eigenvalue $1$. In the interacting limit, the eigenvalues $\chi_\phi$ are no longer integer~\cite{PhysRevB.95.115106,Debertolis2024,PhysRevResearch.6.023178,10.21468/SciPostPhysCore.6.2.030}.
A correlation matrix $\Xi$ featuring a gap
in its eigenvalues $\chi_\phi$ can
be used to characterize the topology of the many-body ground state. Specifically,
the geometric phase of the correlation matrix allows to characterize the topological classification
of the ground state, and the classification becomes equivalent to the one stemming from the
Bloch Hamiltonian in the non-interacting limit. Since the
many-body ground state of the Kondo lattice does not break time-reversal
symmetry, we define the spinless correlation matrix as
\begin{equation}
\bar \Xi_{ij}(\phi) = \frac{1}{2}\sum_{s} \langle \Omega_\phi | c^\dagger_{is} c_{js} | \Omega_\phi \rangle.
\end{equation}
We denote the eigenstates and eigenvalues of $\bar \Xi_{ij}(\phi)$ as $\bar \chi_\phi$ and $|\bar v_\phi \rangle$. The many-body geometric phase is defined as 
\begin{equation}
\Phi = i\int_{\phi=0}^{2\pi} \sum_{\bar \chi_\phi > \Delta}  \langle \bar v_{\phi} | \partial_{\phi} | \bar v_{\phi} \rangle d \phi,
\label{eq:phi}
\end{equation}
where $\Delta$ denotes the
location of the spectral gap in the entanglement spectrum that we take as $\Delta=0.5$~\footnote{$\Delta$
can be taken as any value inside the correlation gap.}. For a non-interacting
spinful dimerized model, the geometric phase $\Phi$ is equivalent to the Zak phase of each spin sector $0,\pi$
for the trivial and topological configurations. In the presence of a gap in the correlation spectra,
the geometric phase $\Phi$ must remain quantized, and as a result, it allows to characterize
the topological states of a many-body Hamiltonian such as the Kondo lattice model.
\eqref{eq:phi} can be directly computed using the many-body ground state computed exactly for the Kondo
lattice model with tensor networks, which yields a value $\Phi=\pm \pi$
for any non-zero Kondo coupling $J_K>0$. It is finally worth noting that the formulation
of a topological invariant in terms of the pumping of the correlation matrix can
be readily extended to other interacting fermionic states, including interacting
Chern insulators, quantum spin hall insulators, and topological crystalline phases.

Finally, let us comment on the experimental realization of our proposal. 
The one-dimensional electron gas can be realized in van der Waals materials, using
twin boundaries in monolayers~\cite{Zhu2022}, stacking domain walls 
bilayers~\cite{Li2024} or helical networks in twisted bilayers~\cite{Rickhaus2018}. 
The Kondo lattice can be formed by 
depositing single magnetic atoms with scanning tunneling microscopy (STM),
which has allowed to create controllable Kondo systems~\cite{Manoharan2000,PhysRevB.97.035417,PhysRevB.109.195415},
and where the many-body spectral functions is measured through tunneling spectroscopy.

{\it{Conclusion.}} We have shown the emergence of topological zero modes in an
engineered one-dimensional Kondo lattice, 
where the topology is solely driven by Kondo physics.
This phenomenology was first demonstrated with an exact solution of the Kondo lattice model with tensor networks,
and afterward by mapping the Kondo lattice model to an effective topological non-Hermitian model.
Finally, we have introduced a correlation matrix pumping method and showed that it allows us to compute the many-body topological invariant of the Kondo lattice model from the exact wavefunction.
Our results put forward the engineering of Kondo lattices as a promising approach
to create topological zero modes, 
enabling the use of quantum magnetism as a driving force to create
many-body topological phases of matter.

\textbf{Acknowledgements}
We acknowledge the computational resources provided by the Aalto Science-IT project. G.C. is supported by European Union's Horizon 2023 research and innovation programme under the Marie Skłodowska-Curie grant agreement No. 101146565, Z.L., E.L.P., and J.L.L. acknowledge financial support from the Academy of Finland Projects Nos.331342 and 358088, the Nokia Industrial Doctoral School in Quantum Technology, the Jane and Aatos Erkko Foundation and the Finnish Quantum Flagship. 

%

\end{document}




\title{Supplemental Material for ``Topological zero modes and correlation pumping in an engineered Kondo lattice"}

\date{\today}

\maketitle

\section{Effective Hamiltonian analysis}
Here we show the effective Hamiltonian analysis for
\beq 
\begin{aligned}
H_1=&t\sum_{n=1}^{N/4-1}\sum_s\mleft(c_{4n+1,s}^\dag c_{4n,s}+c_{4n,s}^\dag c_{4n+1,s}\mright)\\
H_2=&t\sum_{n=1}^{N/4-1}\sum_s\mleft(c_{4n-2,s}^\dag c_{4n-3,s}+c_{4n-3,s}^\dag c_{4n-2,s}\mright)\\&+\gamma_K\sum_{n\in n_K}\sum_{s}\mleft(c_{n,s}^\dag f_{n,s}+f_{n,s}^\dag c_{n,s}\mright)\\&-i\Gamma\sum_{n\in n_K}\sum_{s}f_{n,s}^\dag f_{n,s}\\
H_{12}=&t\sum_{n=1}^{N/4-1}\sum_s\mleft(c_{4n-2,s}^\dag c_{4n-3,s}+c_{4n-1,s}^\dag c_{4n,s}\mright).
\end{aligned}
\eeq
whose effective Hamiltonian is $H_\text{eff}(\omega)=H_1+\Sigma_e(\omega)$ where $\Sigma_e(\omega)=H_{12}^\dag(\omega-H_2)^{-1}H_{12}$.

We first note that $H_{12}$ is block-diagonal with a periodicity of 4 sites. Thus, solving the self-energy $\Sigma_e(\omega)$ reduces to solving it in a single unit cell with
\beq
\begin{aligned}
H_2=&t\sum_s\mleft(c_{2,s}^\dag c_{3,s}+c_{3,s}^\dag c_{2,s}\mright)\\&+\gamma_K\sum_{n=2,3}\sum_{s}\mleft(c_{n,s}^\dag f_{n,s}+f_{n,s}^\dag c_{n,s}\mright)\\&-i\Gamma\sum_{n=2,3}\sum_{s}f_{n,s}^\dag f_{n,s}\\
H_{12}=&t\sum_s\mleft(c_{2,s}^\dag c_{1,s}+c_{3,s}^\dag c_{4,s}\mright)
\end{aligned}
\eeq
In the matrix form:
\beq
\begin{aligned}
\omega-H_2=&\mleft(\begin{array}{cccc}\omega&-t&-\gamma_K\\-t&\omega&&-\gamma_K\\-\gamma_K&&\omega+i\Gamma\\&-\gamma_K&&\omega+i\Gamma\end{array}\mright)\\
H_{12}=&\mleft(\begin{array}{cc}-t&0\\0&-t\\0&0 \\0&0 \end{array}\mright)
\end{aligned}
\eeq
To get $\Sigma_e(\omega)$, we only need to solve the first two columns of $(\omega-H_2)^{-1}$ due to the form of $H_{12}$:
\beq
(\omega-H_2)^{-1}=\mleft(\begin{array}{cccc}a(\omega+i\Gamma)&b(\omega+i\Gamma)&\cdots&\cdots\\b(\omega+i\Gamma)&a(\omega+i\Gamma)&\cdots&\cdots\\a\gamma_K&b\gamma_K&\cdots&\cdots\\b\gamma_K&a\gamma_K&\cdots&\cdots\end{array}\mright)\\
\eeq
where
\begin{align}
a&=-\frac{\mleft[\omega(\omega+i\Gamma)-\gamma_k^2\mright]}{(\omega+i\Gamma)^2t^2-\mleft[\omega(\omega+i\Gamma)-\gamma_k^2\mright]^2}\\
b&=-\frac{(\omega+i\Gamma)t}{(\omega+i\Gamma)^2t^2-\mleft[\omega(\omega+i\Gamma)-\gamma_k^2\mright]^2}.
\end{align}
Thus,
\beq
\Sigma_e(\omega)=\mleft(\begin{array}{cc}\Delta(\omega)&t'(\omega)\\t'(\omega)&\Delta(\omega)\end{array}\mright)
\eeq
with 
\begin{align}
t'(\omega)&=a(\omega+i\Gamma)t^2\\
\Delta(\omega)&=b(\omega+i\Gamma)t^2.
\end{align}
This gives rise to the effective model Eq. (7) in the main text.

\section{Spin rotation pumping}
Here we present an alternative method for computing the topological invariant of our Kondo lattice model via spin rotation pumping. The effective SSH model Eq.(9) motivates the Zak phase, which is $0$ and $\pi$ for a topologically trivial and non-trivial SSH model, as the topological invariant for our Kondo lattice model. However, due to spin-degeneracy, the Zak phase takes a value of $2\pi$ when our model is topologically non-trivial. In order to overcome this limitation of the Zak phase,
we thus define the topological invariant through a many-body spin rotation pumping that leads to
a well defined topological phase of $\pi$ for a spinful system. 
This is done by considering the spin-rotational twisted boundary condition:
\beq \label{eq_spin_twist}
\begin{aligned}
\mathcal{H}_\text{spin}(\phi) =&t\sum_{s,n=1}^{N-1}\mleft(c_{n+1,s}^\dag c_{n,s}+h.c.\mright)
\\&+\sum_{s,s'} \mleft(e^{\frac{i}{2} \phi \vec{\sigma}_{ss'} \cdot \hat e }c^\dag_{N,s}c_{1,s'}+h.c.\mright)
\\&+J_K\sum_{\substack{s,s',\alpha \\ n\in n_K}}c_{n,s}^\dag\sigma^\alpha_{ss'}c_{n,s'}S_{n}^\alpha
\end{aligned}
\eeq
where $\hat e$ denotes the spin rotation axis. Due to the rotational symmetry of the
Hamiltonanian, any choice of $\hat e$ is equivalent, and for the sake of concreteness
we take $\hat e = \hat z$.
The Hamiltonian has a twist-dependent ground state $|\Omega_\phi\rangle$, allowing to define the following geometric phase as the topological invariant
\beq
\Phi_\text{spin}=i\int_{\phi=0}^{2\pi} \langle \Omega_\phi | \partial_{\phi} | \Omega_\phi \rangle d \phi.
\eeq
The geometric phase takes a value of $\pm\pi$ for $J_K>0$, demonstrating the topological origin of the zero modes.
